\documentstyle[12pt]{article}

\newcommand{\states}{Q}
\newcommand{\dirs}{\{L,R\}}

\newcommand{\C}{{\bf C}}

\newcommand{\qed}{\hspace*{\fill}$\Box$ \bigskip}
\newcommand{\blank} {{\#}}
\newcommand{\col}[1] {{| #1 \rangle}}
\newcommand{\iprod}[2] {{\langle #1 | #2 \rangle}}
\newcommand{\up} {{\delta}}
\newcommand{\len}[1] {{\left| #1 \right|}}
\newcommand{\abs}[1] {{\left| #1 \right|}}
\newcommand{\card}[1] {{\mbox{card}(#1)}}
\newcommand{\strlen}[1] {{|#1|}}
\newcommand{\quest} {q_q}
\newcommand{\ans} {q_a}

\newcommand{\PT}{\mbox{\bf P}}
\newcommand{\NP}{\mbox{\bf NP}}
\newcommand{\NPs}{\mbox{\scriptsize\bf NP}}
\newcommand{\npc}{\mbox{\NP--com}\-plete}
\newcommand{\coNP}{\mbox{\bf co--NP}}
\newcommand{\BPP}{\mbox{\bf BPP}}
\newcommand{\PSPACE}{\mbox{\bf PSPACE}}
\newcommand{\BQP}{\mbox{\bf BQP}}
\newcommand{\BQT}[1]{\mbox{\bf BQTime}(#1)}
\newcommand{\BQPs}{\mbox{\scriptsize\bf BQP}}

\newcommand{\suivant}{\\[-6.5mm]\pagebreak[1]}
\newenvironment{proof}{
\par
\noindent {\bf Proof.}\rm}%
{\qed}

\begin{document}

\newtheorem{theorem}{Theorem}[section]
\newtheorem{corollary}[theorem]{Corollary}
\newtheorem{lemma}[theorem]{Lemma}
\newtheorem{fact}[theorem]{Fact}
\newtheorem{definition}[theorem]{Definition}

\title{Strengths and Weaknesses of Quantum Computing}

\author{
Charles H. Bennett \\ {\protect\small\sl IBM Research\/}\,%
\thanks{\,IBM T.\,J. Watson Research Laboratory, Yorktown Heights,
New York, NY 10598, USA.\hfill\mbox{} \mbox{email: bennetc@watson.ibm.com}.}
\\[-1ex] \hspace{5cm}
\and
Ethan Bernstein \\ {\protect\small\sl Microsoft Corporation\/}\,%
\thanks{\,1~Microsoft Way, Redmond, \mbox{WA 98052\,--\,6399}, USA.
\mbox{email: ethanb@microsoft.com}.}
\\[-1ex] \hspace{5cm}
\and
Gilles Brassard\,%
\thanks{\,Supported in part by Canada's {\sc nserc} and Qu\'ebec's
{\sc fcar}.}\\
{\protect\small\sl Universit\'e de Montr\'eal\/}\,%
\thanks{\,D\'epartement IRO, Universit\'e de Montr\'eal,
C.P. 6128, succursale centre-ville,\hfill\mbox{}
\mbox{Montr\'eal (Qu\'ebec), Canada H3C 3J7.
email: brassard@iro.umontreal.ca}.}
\\[-1ex] \hspace{5cm}
\and
Umesh Vazirani\,%
\thanks{\,Supported by NSF Grant No.~CCR-9310214.}
\\ {\protect \small \sl UC Berkeley\/}\,%
\thanks{\,Computer Science Division, University of California, Berkeley,
\mbox{CA 94720}, USA.\hfill\mbox{}
\mbox{email: vazirani@cs.berkeley.edu}.}
\\[-1ex] \hspace{5cm}
}

\date{12 December 1996}

\maketitle

\begin{abstract}

Recently a great deal of attention has focused on quantum computation
following a sequence of results~\cite{BV,Si,Sh} suggesting that
quantum computers are more powerful than classical probabilistic
computers. Following Shor's result that factoring and
the extraction of discrete logarithms
are both solvable in quantum polynomial time, it is natural to
ask whether all of $\NP$ can be efficiently solved in quantum polynomial time.
In this paper, we address this question by proving that
relative to an oracle chosen uniformly at random, with probability 1,
the class $\NP$ cannot be solved on a quantum Turing machine in time
$o(2^{n/2})$. We also show that relative to a permutation oracle chosen
uniformly at random, with probability 1, the class $\NP \cap \coNP$ cannot be
solved on a quantum Turing machine in time $o(2^{n/3})$.  The former bound is
tight since recent work of Grover~\cite{Gr} shows how to accept the class $\NP$
relative to any oracle on a quantum computer in time $O(2^{n/2})$.

\end{abstract}
\thispagestyle{empty}

\vspace{-18cm}
\begin{flushright}
To appear in {\sl SIAM Journal on Computing}\\
(special issue on quantum computing)
\end{flushright}

\newpage

\section{Introduction}

Quantum computational complexity is an exciting new area that touches upon
the foundations of both theoretical computer science and quantum physics.
In~the early eighties, Feynman~\cite{Fe} pointed out that straightforward
simulations of quantum mechanics on a classical computer appear to require a
simulation overhead that is exponential in the size of the system and the
simulated time; he asked whether this is inherent, and whether it is possible
to design a universal quantum computer. Deutsch~\cite{De} defined a general
model of quantum computation: the quantum Turing machine.  Bernstein and
Vazirani~\cite{BV} proved that there is an efficient universal quantum Turing
machine. Yao~\cite{Ya} extended this by proving that quantum circuits
(introduced by Deutsch~\cite{De2}) are polynomially equivalent to quantum
Turing machines.

The computational power of quantum Turing machines (QTMs) has been explored
by several researchers.  Early work by Deutsch and Jozsa~\cite{DJ} showed
how to exploit some inherently quantum mechanical features of QTMs.
Their results, in conjunction with subsequent results by
Berthiaume and Brassard~\cite{BB92,BB94}, established
the existence of oracles under which
there are computational problems that QTMs can solve
in polynomial time with certainty, whereas if we require a classical
probabilistic Turing machine to produce the correct answer with
certainty, then it must take exponential time on some inputs.
On the other hand, these computational problems
are in $\BPP$\,\footnote{\samepage\,$\BPP$ is the class of
decision problems (languages)
that can be solved in polynomial time by probabilistic Turing machines
with error probability bounded by~1/3 (for all inputs).  Using standard
boosting techniques, the error probability can then be made exponentially
small in $k$ by iterating the algorithm $k$ times and returning the majority
answer.} relative to the same oracle, and therefore efficiently
solvable in the classical sense.  The quantum analogue of the
class $\BPP$ is the class $\BQP$\,\footnote{\samepage\,$\BQP$ is the class of
decision problems (languages)
that can be solved in polynomial time by quantum Turing machines
with error probability bounded by~1/3 (for all inputs)---see~\cite{BV}
for a formal definition.
We~prove in Section~\ref{section.subroutine} of this paper that,
as~is the case with $\BPP$, the error probability of $\BQP$ machines
can be made exponentially small.}~\cite{BB92}.
Bernstein and Vazirani~\cite{BV} proved that
$\BPP \subseteq \BQP \subseteq \PSPACE$, thus establishing that it will
not be possible to conclusively prove that $\BQP \neq \BPP$ \mbox{without}
resolving the major open problem $\PT \stackrel{?}{=} \PSPACE$.
They also gave the first evidence that $\BQP \neq \BPP$ (polynomial-time
quantum Turing machines are more powerful than polynomial-time probabilistic
Turing machines), by proving the existence of an oracle relative to which
there are problems in $\BQP$ that cannot be solved with small error
probability by probabilistic machines restricted to running in
$n^{o(\log n)}$ steps.
Since $\BPP$ is regarded as the class of all ``efficiently computable''
languages (computational problems), this provided evidence that quantum
computers are inherently more powerful than classical computers in a
model-independent way.  Simon~\cite{Si} strengthened this evidence by proving
the existence of an oracle relative to which $\BQP$ cannot even be simulated
by probabilistic machines allowed to run for $2^{n/2}$ steps.
In addition, Simon's paper also introduced an important new technique which
was one of the ingredients in a remarkable result proved subsequently by
Shor~\cite{Sh}.  Shor gave polynomial-time quantum algorithms for the factoring
and discrete logarithm problems.  These two problems have been well-studied,
and their presumed \mbox{intractability} forms the basis of much of modern
cryptography.  In view of these results, it is natural to ask whether
$\NP \subseteq \BQP$; i.e.~can quantum computers solve \npc{} problems in
polynomial time?\,\footnote{\samepage\,Actually it is not even clear whether
$\BQP \subseteq \BPP^{\NPs}$; i.e.~it is unclear whether
nondeterminism together with randomness is sufficient to simulate quantum
Turing machines. In~fact, Bernstein and Vazirani's~\cite{BV} result is stronger
than stated above.  They actually proved that relative to an oracle, the
recursive Fourier sampling problem can be solved in $\BQP$, but cannot even be
solved by Arthur-Merlin games~\cite{babai} with a time bound of $n^{o(\log n)}$
(thus giving evidence that nondeterminism on top of probabilism
does not help).  They conjecture that the recursive Fourier sampling cannot
even be solved in the unrelativized polynomial-time hierarchy.}

In this paper, we address this question by proving that
relative to an oracle chosen uniformly at random~\cite{BG}, with probability 1,
the class $\NP$ cannot be solved on a quantum Turing machine in time
$o(2^{n/2})$. We also show that relative to a permutation oracle chosen
uniformly at random, with probability 1, the class $\NP \cap \coNP$ cannot be
solved on a quantum Turing machine in time $o(2^{n/3})$.  The former bound is
tight since recent work of Grover~\cite{Gr} shows how to accept the class $\NP$
relative to any oracle on a quantum computer in time $O(2^{n/2})$.
See~\cite{BBHT} for a detailed analysis of Grover's algorithm.

What is the relevance of these oracle results?
We should emphasize that they do not rule out the possibility
that $\NP \subseteq \BQP$.
What these results do establish is that
there is no black-box approach to solving \npc{} problems by using some
uniquely quantum-mechanical features of QTMs. That this was a real possibility
is clear from Grover's~\cite{Gr} result, which gives a black-box approach to
solving \npc{} problems in square-root as much time as is required classically.

One way to think of an oracle is as a special subroutine call whose
invocation only costs unit time. In the context of QTMs, subroutine
calls pose a special problem that has no classical counterpart. The
problem is that the
subroutine must not leave around any bits beyond its computed answer,
because otherwise computational paths with different residual information
do not interfere.
This is easily achieved for deterministic
subroutines since any classical deterministic computation can be
carried out reversibly so that only the input and the answer remain.
However, this leaves open the more general question of whether a
$\BQP$ machine can be used as a subroutine.  Our final result in this
paper is to show how any $\BQP$ machine can be modified into a {\em tidy\/}
$\BQP$ machine whose final superposition consists almost entirely of a
tape configuration containing just the input and the single bit
answer.  Since these tidy $\BQP$ machines can be safely used as
subroutines, this allows us to show that $\BQP^{\BQPs} = \BQP$. The result
also justifies the definition of oracle quantum machines that we now give.

\section{Oracle Quantum Turing Machines}
\label{section.oracle}

In this section and the next,
we shall assume without loss of generality that the
Turing machine alphabet (for each track or tape) is \mbox{$\{0,1,\blank\}$},
where ``$\blank$'' denotes the blank symbol.  Initially all tapes are blank
except that the input tape contains the actual input surrounded by blanks.
We~shall use $\Sigma$ to denote \mbox{$\{0,1\}$}.

In the classical setting, an oracle may be described informally as a device
for evaluating some Boolean function $A: \Sigma^* \rightarrow \Sigma$, on
arbitrary arguments, at unit cost per evaluation.
This allows to formulate questions such as
``if~$A$ were efficiently computable by a Turing machine,
which other functions (or languages) could be efficiently
computed by Turing machines?''.  In~the quantum setting, an equivalent
question can be asked, provided we define oracle quantum
Turing machines appropriately---which we do in this section---and
provided bounded-error quantum Turing machines can be composed---which
we show in Section~\ref{section.subroutine} of this paper.

An oracle QTM has a special {\em query tape\/} (or track), all of
whose cells are blank except for a single block of non-blank cells.
In a well-formed oracle QTM, the Turing machine rules may allow this region
to grow and shrink, but prevent it from
fragmenting into non-contiguous blocks.\,\footnote{\,This restriction
can be made without loss of generality and it can be verified
syntactically by allowing only machines that make sure they
do not break the rule before writing on the query tape.}
Oracle QTMs have two distinguished internal states: a pre-query state~$\quest$
and a post-query state~$\ans$.  A~query is executed whenever
the machine enters the pre-query state.
If~the query string is empty, a no-op occurs, and the machine passes directly
to the post-query state with no change.  If the query string is nonempty, it
can be written in the form $x \circ b$ where $x \in \Sigma^*$,
$b \in \Sigma$, and ``$\circ$'' denotes concatenation.
In~that case, the result of a call on oracle $A$ is that internal
control passes to the post-query state while the contents of the query tape
changes from $|x\circ b\rangle$ to $|x \circ (b\oplus A(x))\rangle$,
where ``$\oplus$'' denotes the exclusive-or (addition modulo~2).
Except for the query tape and internal
control, other parts of the oracle QTM do not change during the query.
If the target bit $|b\rangle$ is supplied in initial state~$|0\rangle$,
then its final state will be $|A(x)\rangle$, just as in a classical
oracle machine.  Conversely, if the target bit is already in state
$|A(x)\rangle$, calling the oracle will reset it to $|0\rangle$. This ability
to ``uncompute'' will often prove essential to allow proper
interference among computation paths to take place.
Using this fact, it is also easy to see that
the above definition of oracle Turing machines yields unitary evolutions
if we restrict ourselves to
machines that are well-formed in other respects, in particular evolving
unitarily as they enter the pre-query state and leave the post-query state.

The power of quantum computers comes from their ability to follow a
coherent superposition of computation paths.  Similarly oracle quantum
machines derive great power from the ability to perform superpositions
of queries.  For example, oracle $A$ might be called
when the query tape is in state
$|\psi\circ 0\rangle=\sum_x \alpha_x|x\circ 0\rangle$,
where $\alpha_x$ are complex coefficients, corresponding to an arbitrary
superposition of queries with a constant $|0\rangle$ in the target bit.
In~this case, after the query, the query string will be left in the entangled
state $\sum_x \alpha_x | x \circ A(x) \rangle$.
It~is also useful to be able to put the target bit $b$ into a
superposition. For example, the conditional phase inversion used in
Grover's algorithm can be achieved by performing queries with the target
bit $b$ in the nonclassical superposition
$\beta=(|0\rangle-|1\rangle)/\sqrt{2}$. It can readily be verified
that an oracle call with the query tape in state $x\circ\beta$ leaves
the entire machine state, \mbox{including} the query tape, unchanged if
$A(x)=0$, and leaves the entire state unchanged while introducing a
phase factor $-1$ if $A(x)=1$.

It is often convenient to think of a Boolean oracle as defining
a length-preserving function on $\Sigma^*$.
This is easily accomplished by interpreting the oracle answer
on the pair $(x, i)$ as the $i^{th}$ bit of the function value.
The pair $(x, i)$ is encoded as a binary string using any
standard pairing function.  A~{\em permutation oracle\/} is
an oracle which, when interpreted as a length-preserving function, acts
for each $n \geq 0$ as a permutation on $\Sigma^n$.
Henceforth, when no confusion may arise, we shall use $A(x)$ to denote the
length-preserving function associated with oracle $A$
rather than the Boolean function that gives rise to~it.

Let us define $\BQT{T(n)}^A$ as the sets of languages accepted
with probability at least $2/3$ by some oracle QTM $M^A$ whose
running time is bounded by~$T(n)$.
This bound on the running time applies to each individual input,
not just on the average.
Notice that whether or not $M^A$ is
a $\BQP$-machine might depend upon the oracle~$A$---thus $M^A$ might
be a $\BQP$-machine while $M^{B}$ might not be one.

\noindent
{\bf Note:}
The above definition of a quantum oracle for an arbitrary Boolean function
will suffice for the purposes of the present paper, but the ability of
quantum computers to perform general unitary transformations suggests a
broader definition, which may be useful in other contexts.
For~example, oracles that perform more general, non-Boolean unitary
operations have been considered in computational learning
theory~\cite{COLT} and for hiding information against classical
queries~\cite{machta}.

Most broadly, a quantum oracle may be defined as a device that,
when called, applies a fixed unitary transformation $U$ to the
current contents $|z\rangle$ of the query tape, replacing it
by~$U|z\rangle$.  Such an oracle $U$ must be
defined on a countably infinite-dimensional Hilbert space, such as that
spanned by the binary basis vectors $|\epsilon\rangle, |0\rangle, |1\rangle,
|00\rangle, |01\rangle, |10\rangle, |11\rangle, |000\rangle$,~\ldots, where
$\epsilon$ denotes the empty string.
Clearly, the use of such general unitary oracles still yields unitary
evolution for well-formed oracle Turing machines.
Naturally, these oracles
can map inputs onto superpositions of outputs, and vice
versa, and they need not even be length-preserving.  However, in order to
obey the dictum that a single machine cycle ought not to make
infinite changes in the tape, one might require that
$U|z\rangle$ have amplitude zero on all but finitely many basis vectors.
(One~could even insist on a uniform and effective version of the
above restriction.)
Another natural restriction one may wish to impose upon $U$ is that
it be an involution, \mbox{$U^2=I$}, so that the effect of an oracle
call can be undone by a further call on the same oracle.  Again this
may be crucial to allow proper interference to take place.
Note that the special case of unitary transformation considered in this
paper, which corresponds to evaluating a classical Boolean function,
is an involution.

\section{Difficulty of Simulating Nondeterminism on QTMs}
\label{section.rand.oracle}

The computational power of QTMs lies in their ability to
maintain and compute with expo\-nen\-tially large superpositions.
It is tempting to try to use this ``exponential parallelism''
to simulate non-determinism. However, there are inherent
constraints on the scope of this parallelism,
which are imposed by the formalism of quantum mechanics.\,\footnote{\,There
is a superficial similarity between this exponential parallelism
in quantum computation and the fact that probabilistic computations
yield probability distributions over exponentially large domains.
The~difference is that in the
probabilistic case, the computational path is chosen by making
a sequence of random choices---one for each step. In the quantum-mechanical
case, it is possible for several
computational paths to interfere destructively, and therefore it is
necessary to keep track of the entire superposition at each step to
accurately simulate the system.}
In this section, we explore some of these constraints.

To see why quantum interference can speed up $\NP$ problems
quadratically but not exponentially, consider the problem of
distinguishing the empty oracle ($\forall_x A(x)\!=\!0$) from an oracle
containing a single random unknown string $y$ of known length~$n$
(i.e.~$A(y)\!=\!1$, but $\forall_{x\neq y} A(x)\!=\!0$).  We require that the
computer never answer yes on an empty oracle, and seek to maximize its
``success probability'' of answering yes on a
nonempty oracle.  A~classical computer can do no better than to query
distinct \mbox{$n$--bit} strings at random, giving a success probability
$1/2^n$ after one query and $k/2^n$ after $k$ queries.  How can a
quantum computer do better, while respecting the rule that its overall
evolution be unitary, and, in a computation with a nonempty oracle, all
computation paths querying empty locations evolve exactly as they would
for an empty oracle?  A direct quantum analog of the classical algorithm
would start in an equally-weighted superposition of $2^n$ computation
paths, query a different string on each path, and finally collapse the
superposition by asking whether the query had found the nonempty
location.  This yields a success probability $1/2^n$, the same as the
classical computer.  However, this is not the best way to exploit
quantum parallelism.  Our goal should be to maximize the separation
between the state vector $|\psi_k\rangle$ after $k$ interactions with an
empty oracle, and the state vector $|\psi_k(y)\rangle$ after $k$
interactions with an oracle nonempty at an unknown location $y$.
Starting with a uniform superposition
 \[
 |\psi_0\rangle=\frac{1}{\sqrt{2^n}}\sum_x |x\rangle,
 \]
it is easily seen that the separation after one query is maximized by a
unitary evolution to
 \[
 |\psi_1(y)\rangle
 = \frac{1}{\sqrt{2^n}}\sum_x (-1)^{\delta_{x,y}}|x\rangle
 = |\psi_0\rangle-\frac{2}{\sqrt{2^n}}|y\rangle .
 \]
This is a phase inversion of the term corresponding to the nonempty
location. By testing whether the post-query state agrees with
$|{\psi_0}\rangle$ we obtain a success probability
 \[ 1-|\langle\psi_0|\psi_1(y)\rangle|^2 \approx 4/2^n
 \] approximately four times better than the classical value.  Thus, if
we are allowed only one query, quantum parallelism gives a modest
improvement, but is still overwhelmingly likely to fail because the
state vector after interaction with a nonempty oracle is almost the same
as after interaction with an empty oracle. The only way of producing a
large difference after one query would be to concentrate much of the
initial superposition in the $y$ term before the query, which cannot
be done because that location is unknown.

Having achieved the maximum separation after one query, how best can
that separation be increased by subsequent queries?  Various strategies
can be imagined, but a good one (called ``inversion about the average''
by Grover~\cite{Gr})
is to perform an oracle-independent unitary transformation so as to
change the phase difference into an amplitude difference, leaving the
$y$ term with the same sign as all the other terms but a magnitude
approximately threefold larger.  Subsequent phase-inverting
interactions with the oracle, alternating with oracle-independent
phase-to-amplitude conversions, cause the distance between
$|\psi_0\rangle$ and $|\psi_k(y)\rangle$ to grow linearly with $k$,
approximately as $2k/\sqrt{2^n}$ when \mbox{$k \le \sqrt{N}/2$}.
This results in a quadratic growth of the success probability,
approximately as $4k^2/2^n$ for small~$k$.  The proof
of Theorem 3.5 shows that this approach is
essentially optimal: no quantum algorithm can gain more than
this quadratic factor in success probability compared to classical algorithms,
when attempting to answer $\NP$-type questions formulated
relative to a random oracle.

\subsection{Lower Bounds on Quantum Search}

We will sometimes find it convenient to measure the accuracy of a simulation
by calcu\-lating the Euclidean distance\,\footnote{\,The Euclidean distance
between $|\phi\rangle = \sum_x \alpha_x \col{x}$ and
$|\psi\rangle = \sum_x \beta \col{x}$ is defined as
\mbox{$(\sum_x |\alpha_x - \beta_x|^2)^{1/2}$}.\phantom{\Large X}}
between the target and simulation
superpositions.  The~following theorem from~\cite{BV} shows that the simulation
accuracy is at most $4$ times worse than this Euclidean distance.

\begin{theorem}
\label{total.var.dist}
If two unit-length superpositions
are within Euclidean distance $\varepsilon$ then
observing the two superpositions gives samples from distributions which
are within total variation distance\,\footnote{\,The total variation distance
between two distributions $\cal{D}$ and $\cal{D'}$ is
$\sum_x |{\cal D}(x) -{\cal D'}(x)|$.\phantom{\Large X}}
at most $4 \varepsilon$.
\end{theorem}

\begin{definition}
Let $\col{\phi_i}$ be the superposition of $M^A$ on input $x$
at time $i$. We denote by $q_y (\col{\phi_i})$ the sum of
squared magnitudes in
$\col{\phi_i}$ of configurations of $M$ which are querying
the oracle on string $y$.
We refer to $q_y (\col{\phi_i})$ as the
{\em query magnitude of $y$\/} in~$\col{\phi_i}$.
\end{definition}

\begin{theorem}
\label{flip.answers}
Let $\col{\phi_i}$ be the superposition of $M^A$ on input $x$
at time $i$. Let $\varepsilon > 0$.
Let~\mbox{$F \subseteq [0,T-1] \times \Sigma^*$} be a set of time-strings
pairs such that $\sum_{(i,y) \in F} q_y (\col{\phi_i}) \leq
{\varepsilon^2 \over T}$.  Now~suppose the answer to each query
$(i,y) \in F$ is modified to some arbitrary fixed $a_{i,y}$ (these~answers
need not be consistent with an oracle). Let $\col{\phi '_i}$ be the
time $i$ superposition of $M$ on input $x$ with oracle $A$ modified
as stated above. Then $\len{\col{\phi_T} - \col{\phi '_T}} \leq \varepsilon$.
\end{theorem}

\begin{proof}
Let $U$ be the unitary time evolution operator of $M^A$.
Let $A_i$ denote an oracle such that if $(i,y) \in F$
then $A_i(y) = a_{i,y}$ and if $(i,y) \notin F$ then $A_i (y) = A(y)$.
Let $U_i$ be the unitary time evolution operator of $M^{A_i}$.
Let $\col{\phi_i}$ be the superposition of $M^A$ on input $x$
at time $i$. We define $\col{E_i}$ to be the error in the
$i^{th}$ step caused by replacing the oracle $A$ with~$A_i$. Then
\[\col{E_i} = U_i \col{\phi_{i}} - U \col{\phi_{i}}.\]
So we have
\[\col{\phi_T} = U \col{\phi_{T-1}} = U_T \col{\phi_{T-1}} - \col{E_{T-1}}
= \cdots = U_T \cdots U_1 \col{\phi_0} - \sum_{i = 0}^{T-1} U_{T-1} \cdots
U_{i}\col{E_i}.\]
Since all of the $U_i$ are unitary,
$\len{U_{T-1} \cdots U_{i}\col{E_i}} = \len{\col{E_i}}$.

The sum of squared magnitudes of all of
the $E_i$ is equal to $\sum_{(i,y) \in F} q_y (\col{\phi_i})$
and therefore at most ${\varepsilon^2 \over T^2}$.
In the worst case, the $U_{T-1} \cdots U_{i}\col{E_i}$s
could interfere constructively; however, the squared magnitude
of their sum is at most $T$ times the sum
of their squared magnitudes, i.e.~$\varepsilon ^2$. Therefore
$\len{\col{\phi_T} - \col{\phi '_T}} \leq \varepsilon$.
\end{proof}

\begin{corollary}
\label{change.oracles}
Let $A$ be an oracle over alphabet $\Sigma$. For $y \in \Sigma^*$,
let $A_y$ be any oracle such that $\forall x \neq y ~ A_y (x) = A(x)$.
Let $\col{\phi_i}$ be the time $i$ superposition of $M^A$ on input
$x$ and $\col{\phi_i}^{(y)}$ be the time $i$ superposition of $M^{A_y}$
on input $x$. Then for every $\varepsilon > 0$,
there is a set $S$ of cardinality at most
${2 T^2 \over \varepsilon^2}$ such that
$\forall y \notin S ~ \len{\col{\phi_T}-\col{\phi_T}^{(y)}} \leq \varepsilon$.
\end{corollary}

\begin{proof}
Since each $\col{\phi_t}$ has unit length,
$\sum_{i=0}^{T-1} \sum_{y} q_y (\col{\phi_i}) \leq T$.
Let $S$ be the set of strings $y$ such that
$\sum_{i=0}^{T-1} ~ q_y (\col{\phi_i}) \geq {\varepsilon^2 \over 2 T}$.
Clearly $\card{S} \leq {2 T^2 \over \varepsilon^2}$.

If $y \notin S$ then
$\sum_{i=0}^{T-1} ~ q_y (\col{\phi_i}) < {\varepsilon^2 \over 2 T}$.
Therefore by Theorem~\ref{flip.answers}
$\forall y \notin S ~ \len{\col{\phi_i}-\col{\phi_i}^{(y)}} \leq \varepsilon$.
\end{proof}

\begin{theorem}\label{notNP}
For any $T(n)$ which is $o(2^{n/2})$, relative to a random oracle,
with probability~$1$, $\BQT{T(n)}$ does not contain $\NP$.
\end{theorem}

\begin{proof}
Recall from Section~\ref{section.oracle} that an oracle can
be thought of as a length-preserving function: this is what
we mean below by $A(x)$.
Let ${\cal L}_A = \{y: \exists x ~ A(x) = y\}$.
Clearly, this language is contained in $\NP^A$.
Let $T(n) = o(2^{n/2})$. We show that for any
bounded-error oracle QTM $M^A$ running in time at most $T(n)$,
with probability 1, $M^A$ does not accept the language ${\cal L}_A$.
The probability is taken over the choice of a random length-preserving
oracle~$A$.
Then, since there are a countable number of QTMs
and the intersection of a countable number of probability 1 events
still has probability 1, we conclude that with probability 1,
no bounded error oracle
QTM accepts ${\cal L}_A$ in time bounded by $T(n)$.

Since $T(n) = o(2^{n/2})$, we can pick $n$ large enough so that
$T(n) \leq {2^{n/2} \over 20}$.
We will show that the probability
that $M$ gives the wrong answer on input $1^n$ is at least
$1/8$ for every way of fixing the oracle answers on
inputs of length not equal to $n$. The probability is taken over
the random choices of the oracle for inputs of length $n$.

Let us fix an arbitrary length-preserving function from strings of lengths
other than $n$ over alphabet $\Sigma$. Let ${\cal C}$
denote the set of oracles consistent with this arbitrary
function. Let~${\cal A}$ be the set of oracles in ${\cal C}$ such that $1^n$
has no inverse (does not belong to ${\cal L}_A$). If~the~oracle
answers to length $n$ strings are chosen uniformly at random,
then the probability that the oracle is in ${\cal A}$ is at least
$1/4$. This is because the probability that $1^n$ has no
inverse is $({2^n - 1 \over
2^n})^{2^n}$ which is at least $1/4$ (for $n$ sufficiently large).
Let ${\cal B}$ be
the set of oracles in ${\cal C}$ such that $1^n$ has a unique inverse.
As above, the probability that a randomly chosen oracle
is in ${\cal B}$ is
$({2^n - 1 \over 2^n})^{2^n-1}$ which is at least $1/e$.

Given an oracle $A$ in ${\cal A}$, we can modify its answer
on any single input, say $y$, to $1^n$ and therefore get an oracle $A_y$
in ${\cal B}$. We will show that for most choices of $y$, the
acceptance probability of $M^A$ on input $1^n$ is almost equal to
the acceptance probability of $M^{A_y}$ on input~$1^n$. On~the other
hand, $M^A$ must reject $1^n$ and $M^{A_y}$ must accept $1^n$. Therefore
$M$ cannot accept both ${\cal L}_A$ and ${\cal L}_{A_y}$. By
working through the details more carefully, it is easy to show that $M$ fails
on input $1^n$ with probability at least $1/8$ when the
oracle is a uniformly random function on strings of length $n$,
and is an arbitrary function on all other strings.

Let $A_y$ be the oracle such that
$A_y (y) = 1^n$ and $\forall z \neq y ~ A_y(z) = A(z)$.
By Corollary~\ref{change.oracles} there is a set $S$ of at most
$338 T^2(n)$ strings such that the difference between the $i^{th}$
superposition of $M^{A_y}$ on input $1^n$ and
$M^A$ on input $1^n$ has norm at most $1/13$.
Using Theorem~\ref{total.var.dist} we can conclude that the difference between
the acceptance probabilities of $M^{A_y}$ on input $1^n$ and
$M^A$ on input $1^n$ is at most $1/13 \times 4 < 1/3$.
Since $M^{A_y}$ should accept $1^n$ with probability at least
$2/3$ and $M^A$ should reject $1^n$ with probability at least
$2/3$, we can conclude that $M$ fails to accept either
${\cal L}_A$ or ${\cal L}_{A_y}$.

So, each oracle $A \in {\cal A}$ for which $M$ correctly decides
whether $1^n \in {\cal L}_A$ can, by changing a single answer
of $A$ to $1^n$, be mapped to at least $(2^n - \card{S}) \geq 2^{n-1}$
different oracles $A_f \in {\cal B}$ for which $M$ fails to correctly
decide whether $1^n \in {\cal L}_{A_f}$.  Moreover, any particular
$A_f \in {\cal B}$ is the image under this mapping of at most $2^n -1$ oracles
$A \in {\cal A}$, since where it now answers $1^n$, it must have given
one of the $2^n - 1$ other possible answers.
Therefore, the number of oracles in ${\cal B}$ for which $M$ fails
must be at least $1/2$ the number of oracles in
${\cal A}$ for which $M$ succeeds.
So, calling $a$ the number of oracles in $\cal A$ for which $M$ fails,
$M$ must fail for at least $a + 1/2(\card{\cal A} - a)$ oracles.
Therefore $M$ fails to correctly decide whether $1^n \in {\cal L}_A$
with probability at least $(1/2) P[{\cal A}] \geq 1/8$.

It is easy to conclude that $M$ decides
membership in ${\cal L}_A$ with probability $0$ for
a uniformly chosen oracle $A$.
\end{proof}

\noindent
{\bf Note:}
Theorem~\ref{flip.answers} and its Corollary~\ref{change.oracles}
isolate the constraints on ``quantum parallelism'' imposed by
unitary evolution. The rest of the proof of the above theorem is
similar in spirit to standard techniques
used to separate $\BPP$ from $\NP$ relative to a random oracle~\cite{BG}.
For~example, these techniques can be used to show that, relative to a
random oracle~$A$, no classical
probabilistic machine can recognize ${\cal L}_A$ in time~$o(2^n)$.
However, quantum machines can recognize this language quadratically faster,
in time~$O(\sqrt{2^n}\,)$,
using Grover's algorithm~\cite{Gr}. This explains why a substantial
modification of the standard technique was required to prove the above
theorem.

The next result about $\NP \cap \coNP$ relative to a random
permutation oracle requires a more subtle argument; ideally
we would like to apply Theorem~\ref{flip.answers} after asserting that
the total query magnitude with which $A^{-1}(1^n)$ is
probed is small.  However, this is precisely what we are
trying to prove in the first place.

\begin{theorem}\label{notCoNP}
For any $T(n)$ which is $o(2^{n/3})$, relative to a random permutation
oracle,
with probability $1$, $\BQT{T(n)}$
does not contain $\NP \cap \coNP$.
\end{theorem}

\begin{proof}
For any permutation oracle $A$, let
${\cal L}_A = \{y: \mbox{~first~bit~of~}A^{-1}(y) \mbox{~is~} 1 \}$.
Clearly, this language is contained in $(\NP \cap \coNP)^A$.
Let $T(n) = o(2^{n/3})$. We show that for any
bounded-error oracle QTM $M^A$ running in time at most $T(n)$,
with probability 1, $M^A$ does not accept the language ${\cal L}_A$.
The probability is taken over the choice of a random permutation
oracle $A$.
Then, since there are a countable number of QTMs
and the intersection of a countable number of probability 1 events
still has probability 1, we conclude that with probability 1,
no bounded error oracle
QTM accepts ${\cal L}_A$ in time bounded by $T(n)$.

Since $T(n) = o(2^{n/3})$, we can pick $n$ large enough so that
$T(n) \leq {2^{n/3} \over 100}$.
We will show that the probability
that $M$ gives the wrong answer on input $1^n$ is at least
$1/8$ for every way of fixing the oracle answers on
inputs of length not equal to $n$. The probability is taken over
the random choices of the permutation
oracle for inputs of length $n$.

Consider the following method of defining random permutations
on $\{ 0,1 \}^n$: let $x_0 , x_1 , \ldots x_{T+1}$ be a sequence of
strings chosen uniformly at random in $\{ 0, 1 \}^n$. Pick
$\pi_0$ uniformly at random among permutations such
that $\pi(x_0) = 1^n$. Let $\pi_i = \pi_{i-1} \cdot \tau$,
where $\tau$ is the transposition $(x_{i-1} , x_i )$,
i.e.~$\pi_i (x_i) = \pi_{i-1} (x_{i-1})$ and
$\pi_i (x_{i-1}) = \pi_{i-1} (x_{i})$. Clearly each $\pi_i$ is
a random permutation on $\{ 0,1\}^n$.

Consider a sequence of permutation oracles $A_i$, such that
$A_i (y) = A_j (y)$ if $y \notin \{0,1\}^n$ and
$A_i (y) = \pi_i (y)$ if $y \in \{0,1\}^n$.
Denote by $\col{\phi_i}$ the time $i$
superposition of $M^{A_{T(n)}}$ on input~$1^n$,
and by $\col{\phi^{'}_i}$ the time $i$
superposition of $M^{A_{T(n)-1}}$ on input $1^n$.
By construction, with probability exactly $1/2$,
the string $1^n$ is a member of exactly one of
the two languages $L_{A_{T(n)}}$ and~$L_{A_{T(n)-1}}$.
We will show that
$E[\len{\col{\phi_{T(n)}} - \col{\phi^{'}_{T(n)}}}] \leq 1/50$.
Here the expectation is taken over
the random choice of the oracles. By Markov's bound,
\mbox{$P[\len{\col{\phi_{T(n)}}-\col{\phi^{'}_{T(n)}}}\leq 2/25]\geq3/4$}.
Applying Theorem~\ref{total.var.dist} we conclude that if
$\len{\col{\phi_{T(n)}} - \col{\phi^{'}_{T(n)}}} \leq 2/25$, then
the acceptance probability of $M^{A_{T(n)}}$ and $M^{A_{T(n)-1}}$ differ by
at most $8/25 < 1/3$, and hence either both machines accept input $1^n$ or
both reject that input. Therefore
$M^{A_{T(n)}}$ and $M^{A_{T(n)-1}}$ give the same
answers on input $1^n$ with probability at least $3/4$.
By construction, the probability that the string $1^n$ belongs to
exactly one of the two languages $L_{A_{T(n)}}$ and $L_{A_{T(n)-1}}$
is equal to $P[$first bit of $x_{T(n)-1} \neq$ first bit
of $x_{T(n)}] = 1/2$.
Therefore, we can conclude that with probability at least
$1/4$, either $M^{A_{T(n)}}$ or $M^{A_{T(n)-1}}$ gives
the wrong answer on input $1^n$.
Since each of $A_{T(n)}$ and $A_{T(n)-1}$ are chosen from
the same distribution, we can conclude that
$M^{A_{T(n)}}$ gives the wrong answer on input $1^n$ with probability
at least~$1/8$.

To bound $E[\len{\col{\phi_{T(n)}} - \col{\phi^{'}_{T(n)}}}]$,
we show that $\col{\phi_{T(n)}}$ and
$\col{\phi^{'}_{T(n)}}$ are each close to a certain superposition
$\col{\psi_{T(n)}}$.
To define this superposition, run $M$ on input $1^n$ with a
different oracle on each
step: on step $i$, use $A_i$ to answer the oracle queries.
Denote by $\col{\psi_i}$, the time $i$ superposition that results.
Consider the set of time-string pairs
\mbox{$S = \{(i, x_j ) :~j \geq i,~0 \leq i \leq T\}$}.
It is easily checked that the oracle queries
in the computation described above and those
of $M^{A_{T(n)}}$ and $M^{A_{T(n)+1}}$ differ only on the set $S$.
We claim that the expected query magnitude of any pair in the
set is at most $1/2^n$, since for $j \geq i$, we may
think of $x_j$ as having been randomly chosen during step $j$,
{\em after\/} the superposition of oracle queries to be performed
has already been written on the oracle tape.
Let $\alpha$ be the
sum of the query magnitudes for time-string pairs in $S$.  Then
\[ E[\alpha] \leq \card{S}/2^n =
{T(n)+1 \choose 2} /2^n \leq \frac {T(n)^2}{2^n} \]
for $T(n) \geq 4$.  Let $\varepsilon$ be a random variable such that
$\alpha = \varepsilon^2/2 T(n)$. Then by Theorem~\ref{flip.answers},
$\len{\col{\phi} - \col{\phi_{T(n)}}} \leq \varepsilon$
and $\len{\col{\phi} - \col{\phi^{'}_{T(n)}}} \leq \varepsilon$.
We showed above that
\[E[\varepsilon^2/T(n)] = E[\alpha] \leq \frac {T(n)^2}{2^n}\,.\]
But $E[\varepsilon/\sqrt{2 T(n)}]^2 \leq E[\varepsilon^2/2 T(n)]$.  Therefore
\[ E[\varepsilon] = \sqrt{2 T(n)}E[\varepsilon/\sqrt{2 T(n)}] \leq
\sqrt{2 T(n) E[\varepsilon^2/2 T(n)]} \leq \sqrt{2 T(n) \frac{T(n)^2}{2^n}}
\leq \sqrt{\frac{2}{100^3}} < 1/100 . \]
Therefore $E[\len{\col{\phi} - \col{\phi_{T(n)}}}] \leq E[\varepsilon] < 1/100$
and $E[\len{\col{\phi} - \col{\phi^{'}_{T(n)}}}] \leq E[\varepsilon]
< 1/100$.  It~follows that $E[\len{\col{\phi_{T(n)}} -
\col{\phi^{'}_{T(n)}}}] < 1/50$.

Finally, it is easy to conclude that $M$ decides membership
in ${\cal L}_A$ with probability $0$ for a uniformly random
permutation oracle $A$.
\end{proof}

\noindent
{\bf Note:}
In~view of Grover's algorithm~\cite{Gr}, we know that
the constant ``$1/2$'' in the statement of Theorem~\ref{notNP}
cannot be improved.  On~the other hand, there is no evidence
that the constant ``$1/3$'' in the statement of Theorem~\ref{notCoNP}
is fundamental.  It~may well be that Theorem~\ref{notCoNP}
would still hold (albeit not its current proof) with $1/2$ substituted
for~$1/3$.

\begin{corollary}
Relative to a random permutation oracle, with probability $1$,
there exists a {\em quantum one-way permutation}.
Given the oracle, this permutation can be computed efficiently even
with a classical deterministic machine, yet it requires exponential
time to invert even on a quantum machine.
\end{corollary}

\begin{proof}
Given an arbitrary permutation oracle~$A$
for which $A^{-1}$ can be computed in time $o(2^{n/3})$
on a quantum Turing machine, it is just as easy to decide
${\cal L}_A$ as defined in the proof of Theorem~\ref{notCoNP}.
It~follows from that proof that this happens with probability~0
when $A$ is a uniformly random permutation oracle.
\end{proof}

\section{Using a Bounded-Error QTM as a Subroutine}
\label{section.subroutine}

The notion of a subroutine call or an oracle invocation provides a simple and
useful abstraction in the context of classical computation. Before making
this abstraction in the context of quantum computation, there are some
subtle considerations that must be thought through. For example, if the
subroutine computes the function $f$, we would like to think of an invocation
of the subroutine on the string $x$ as magically writing $f(x)$ in some
designated spot (actually xoring it to ensure unitarity). In the context of
quantum algorithms, this abstraction is only valid if the subroutine cleans up
all traces of its intermediate calculations, and leaves just the final answer
on the tape. This is because if the subroutine is invoked on a superposition of
$x$'s, then different values of $x$ would result in different scratch-work on
the tape, and would prevent these different computational paths from
interfering. Since erasing is not a unitary operation, the scratch-work
cannot, in general, be erased post-facto. In the special case where $f$ can be
efficiently computed deterministically, it is easy to design the subroutine so
that it reversibly erases the scratch-work---simply compute
$f(x)$, copy $f(x)$ into safe storage, and then uncompute $f(x)$ to get rid of
the scratch work~\cite{oldbennett}. However, in the case that $f$ is computed
by a $\BQP$ machine, the situation is more complicated. This is because only
some of the computational paths of the machine lead to the correct
answer~$f(x)$, and therefore if we copy $f(x)$ into safe storage and then
uncompute $f(x)$, computational paths with different values of $f(x)$ will no
longer interfere with each other, and we will not reverse the first phase of
the computation. We show, nonetheless, that if we boost the success probability
of the $\BQP$ machine before copying $f(x)$ into safe storage and uncomputing
$f(x)$, then most of the weight of the final superposition has a clean tape
with only the input $x$ and the answer $f(x)$. Since such tidy $\BQP$ machines
can be safely used as subroutines, this allows us to show that
$\BQP^{\BQPs} = \BQP$. The result also justifies our definition of
oracle quantum machines.

The correctness of the boosting procedure is proved in
Theorems~\ref{boost} and~\ref{bqp.clean}. The~proof follows the same outline as
in the classical case, except that we have to be much more careful in simple
programming constructs such as looping, etc. We therefore borrow the machinery
developed in~\cite{BV} for this purpose, and present the statements of the
relevant lemmas and theorems in the first part of this section. The main new
contribution in this section is in the proofs of Theorems~\ref{boost}
and~\ref{bqp.clean}. The reader may therefore wish to skip directly ahead to
these proofs.

\subsection{Some Programming Primitives for QTMs}

In this subsection, we present several definitions,
lemmas and theorems from~\cite{BV}.

Recall that a QTM $M$ is defined by a triplet
$(\Sigma,\states,\up)$ where:
$\Sigma$ is a finite alphabet with an identified blank symbol $\blank$,
$\states$ is a finite set of states with an identified initial state $q_0$ and
final state $q_f \neq q_0$,
and $\up$, the {\em quantum transition function},
is a function
\[\up\ \ :\ \ \states \ \times\ \Sigma\ \rightarrow\ \tilde{\C}^{\Sigma\ \times
 \ \states\ \times\ \dirs}\]
where $\tilde{\C}$ is the set of complex numbers whose real and
imaginary parts can be approximated to within $2^{-n}$ in time
polynomial in $n$.

\begin{definition} A {\em final configuration\/} of a QTM is any configuration
in state $q_f$.
If when QTM $M$ is run with input $x$, at time
$T$ the superposition contains only final configurations and at any
time less than $T$ the superposition contains no final configuration,
then $M$ {\em halts\/} with {\em running time\/} $T$ on input $x$.
The superposition of $M$ at time $T$ is called the {\em final superposition\/}
of $M$ run on input $x$.  A {polynomial-time\/} QTM is a well-formed
QTM which on every input $x$ halts in time
polynomial in the length of $x$.
\end{definition}

\begin{definition}
A QTM $M$ is called {\em well-behaved\/} if it halts on all input strings in
a final superposition where each configuration has the tape head in the same
cell.  If this cell is always the start cell, we call the QTM {\em stationary}.
\end{definition}

We will say that a QTM $M$ is in {\em normal form\/} if all transitions from
the distinguished state $q_f$ lead to the distinguished state $q_0$,
the symbol in the scanned cell is left unchanged, and the head moves
right, say. Formally:

\begin{definition}
A QTM
$M = (\Sigma, \states, \delta)$
is in {\em normal form\/} if
\[ \forall \sigma \in \Sigma\ \ \ \delta(q_f,\sigma) =
   \col{\sigma}\col{q_0}\col{R} \]
\end{definition}

\begin{theorem}
\label{synchronization}
If $f$ is a function mapping strings to strings
which can be computed in deterministic polynomial time and
such that the length of $f(x)$ depends only on the length of $x$,
then there is a polynomial-time, stationary, normal form QTM
which given input $x$, produces output $x;f(x)$,
and whose running time depends only on the
length~of~$x$.

If $f$ is a one-to-one function from strings to strings that such that both
$f$ and $f^{-1}$ can be computed in deterministic polynomial time,
and such that the length of $f(x)$ depends only on the length of $x$,
then there is a polynomial-time, stationary, normal form QTM
which given input $x$, produces output $f(x)$,
and whose running time depends only on the
length~of~$x$.
\end{theorem}

\begin{definition} A {\em multi-track\/} Turing machine with $k$ tracks
is a Turing machine whose alphabet $\Sigma$ is of the form
$\Sigma_1 \times \Sigma_2 \times \cdots \times \Sigma_k$ with a special
blank symbol $\blank$ in each $\Sigma_i$ so that the blank in $\Sigma$ is
$(\blank,\ldots,\blank)$. We specify the input by specifying the string on each
``track'' (separated by `;'), and optionally by specifying the
alignment of the contents of the tracks.
\end{definition}

\begin{lemma}
\label{extension.lemma}
Given any QTM
$M = (\Sigma,\states,\up)$ and any
set $\Sigma'$, there is a QTM
\mbox{$M' = (\Sigma \times \Sigma',\states,\up')$} such that $M'$
behaves exactly as $M$ while leaving its second track \mbox{unchanged}.
\end{lemma}

\begin{lemma}
\label{permutation.lemma}
Given any QTM
$M = (\Sigma_1 \times \cdots \times \Sigma_k,\states,\up)$
and permutation $\pi : [1,k] \rightarrow [1,k]$, there is a QTM
$M' = (\Sigma_{\pi(1)} \times \cdots \times \Sigma_{\pi(k)},\states,\up')$
such that the $M'$ behaves exactly as $M$ except that its tracks are
permuted according to $\pi$.
\end{lemma}

\begin{lemma}
\label{dovetailing}
If $M_1$ and $M_2$ are well-behaved, normal form QTMs
with the same alphabet, then there is a normal form QTM
$M$ which carries out the computation of $M_1$ followed by the
computation~of~$M_2$.
\end{lemma}

\begin{lemma}
\label{looping}
Suppose that $M$ is a well-behaved, normal form QTM.
Then there is a normal form QTM $M'$ such that on input $x;k$ with
$k > 0$, the machine $M'$ runs $M$ for $k$ iterations on its first track.
\end{lemma}

\begin{definition}
If QTMs $M_1$ and $M_2$ have the same alphabet, then we say that $M_2$
{\em reverses\/} the computation of $M_1$ if the following holds:
for any input $x$ on which $M_1$ halts, let $c_x$ and $\phi_x$ be the initial
configuration and final superposition of $M_1$ on input $x$. Then
$M_2$ on input the superposition $\phi_x$, halts with
final superposition consisting entirely of configuration~$c_x$.
Note that for $M_2$ to reverse $M_1$, the final state of $M_2$ must
be equal to the initial state of $M_1$ and vice versa.
\end{definition}

\begin{lemma}
\label{reverse}
If $M$ is a normal form QTM which halts on all inputs,
then there is a normal form
QTM $M'$ that reverses the computation of $M$ with slowdown by a factor of 5.
\end{lemma}

Finally, recall the definition of the class $\BQP$.

\begin{definition}
Let $M$ be a stationary, normal form, multi-track QTM $M$ whose last track has
alphabet $\{\blank,0,1\}$. We say that $M$ accepts $x$ if it halts with a
$1$ in the last track of the start cell. Otherwise we say that $M$ rejects $x$.

A QTM {\em accepts\/} the language ${\cal L} \subseteq (\Sigma - \blank)^*$
{\em with probability $p$\/} if $M$ accepts with probability at least $p$ every
string $x \in {\cal L}$ and rejects with probability at least $p$ every string
\mbox{$x \in (\Sigma - \blank)^* - {\cal L}$}. We~define the class $\BQP$
{\em (bounded-error quantum polynomial time)\/} as the set of languages
which are accepted with probability $2/3$ by some polynomial-time~QTM.
More generally, we define the class $\BQT{T(n)}$ as the set of
languages which are \mbox{accepted} with probability $2/3$ by some QTM
whose running time on any input of length $n$ is bounded by $T(n)$.
\end{definition}

\subsection{Boosting and Subroutine Calls}
\begin{theorem}
\label{boost}
If QTM $M$ accepts language ${\cal L}$ with probability $2/3$ in time
$T(n) > n$, with $T(n)$ time-constructible,
then for any $\varepsilon > 0$, there is a QTM $M'$ which accepts ${\cal L}$
with probability $1 - \varepsilon$ in time $cT(n)$ where
$c$ is polynomial in $\log 1/\varepsilon$ but independent of~$n$.
\end{theorem}

\begin{proof}
Let $M$ be a stationary QTM which accepts the language ${\cal L}$ in time
$T(n)$.

We will build a machine that runs $k$ independent copies of $M$
and then takes the \mbox{majority} vote of the $k$ answers.
On any input $x$, $M$ will have some final superposition of strings
$\sum_{i} \alpha_i \col{x_i}$.  If we call $A$ the set of $i$
for which $x_i$ has the correct answer $M(x)$
then $\sum_{i \in A} \abs{\alpha_i}^2 \geq 2/3$.
Now running $M$ on separate copies of its input $k$ times will produce
$\sum_{i_1,\ldots,i_k} \alpha_{i_1} \cdots \alpha_{i_k} \col{x_{i_1}} \cdots
\col{x_{i_k}}$. Then the probability of seeing $\col{x_{i_1}} \cdots
\col{x_{i_k}}$ such that the majority have the correct answer $M(x)$ is the sum
of $\abs{\alpha_{i_1}}^2 \cdots \abs{\alpha_{i_k}}^2$ such that
the majority of $i_1,\ldots,i_k$ lie in $A$.  But this is just
like taking the majority of $k$ independent coin flips each with probability
at least $2/3$ of heads.  Therefore there is some constant
$b$ such that when $k = b \log 1/\varepsilon$, the probability
of seeing the correct answer will be at least $1 - \varepsilon$.

So, we will build a machine to carry out the following steps.

\begin{enumerate}
\item Compute $n = T(\strlen{x})$.
\item Write out $k$ copies of the input $x$ spaced out with $2n$ blank
cells in between, and write down $k$ and $n$ on other tracks.
\item Loop $k$ times on a machine that runs $M$ and then steps $n$ times to
the right.
\item Calculate the majority of the $k$ answers and write it back in the
start cell.
\end{enumerate}

We construct the desired QTM by building
a QTM for each of these four steps and then dovetailing them together.

Since Steps 1, 2, and 4 require easily computable functions whose output length
depend only on $k$ and the length of $x$,
we can carry them out using well-behaved, normal form QTMs,
constructed using Theorem~\ref{synchronization}, whose running times
also depend only on $k$ and the length of $x$.

So, we complete the proof by constructing a QTM to run the given machine $k$
times. First, using Theorem~\ref{synchronization} we can construct a
stationary, normal form QTM which drags the integers $k$ and $n$ one square to
the right on its work track. If we add a single step right to the end of this
QTM and apply Lemma~\ref{looping}, we can build a well-behaved, normal form QTM
moves which $n$ squares to the right, dragging $k$ and $n$ along with it.
Dovetailing this machine after $M$, and then applying Lemma~\ref{looping}
gives a normal form QTM that runs $M$ on each of the $k$ copies of the input.
Finally, we can dovetail with a machine to return with $k$ and $n$ to the start
cell by using Lemma~\ref{looping} two more times around a QTM which carries $k$
and $n$ one step to the left.
\end{proof}

The extra information on the output tape of a QTM can be erased by copying
the desired output to another track, and then running the reverse of the
QTM.  If the output is the same in every configuration in the final
superposition, then this reversal will exactly recover the input.
Unfortunately, if the output differs in different configurations, then saving
the output will prevent these configurations from interfering when the machine
is reversed, and the input will not be recovered. We show
is the same in most of the final superposition, then the reversal must
lead us close to the input.

\begin{theorem}
\label{bqp.clean}
If the language ${\cal L}$ is contained in the class $\BQT{T(n)}$, with
$T(n) > n$ and $T(n)$ time-constructible,
then for any $\varepsilon > 0$, there is a QTM $M'$ which accepts ${\cal L}$
with probability $1 - \varepsilon$ and has the following property.
When run on input $x$ of length $n$, $M'$ runs for time bounded by $c T(n)$,
where $c$ is a polynomial in $\log 1/\varepsilon$, and produces a final
superposition in which $\col{x}\col{{\cal L}(x)}$, with ${\cal L}(x) = 1$ if
$x \in {\cal L}$ and $0$ otherwise, has squared magnitude at
least~$1 - \varepsilon$.
\end{theorem}

\begin{proof}
Let $M = (\Sigma,\states,\up)$ be a stationary, normal form QTM which accepts
language ${\cal L}$ in time bounded by $T(n)$.

According to Theorem~\ref{boost}, at the expense of a slowdown by factor
which is polynomial in $\log 1/\varepsilon$ but independent of $n$, we can
assume that $M$ accepts ${\cal L}$ with probability $1 - \varepsilon/2$ on
every input.

Then we can construct the desired $M'$ by running $M$,
copying the answer to another track, and then running the reverse of $M$.
The copy is easily accomplished with a simple two-step machine that steps
left and back right while writing the answer on a clean track.
Using Lemma~\ref{reverse}, we can construct a normal form QTM $M^R$
which reverses $M$.  Finally, with appropriate use
of Lemmas~\ref{extension.lemma} and \ref{permutation.lemma}, we can construct
the desired stationary QTM $M'$ by dovetailing machines $M$ and $M^R$ around
the copying machine.

To see that this $M'$ has the desired properties, consider running $M'$ on
input $x$ of length~$n$.  $M'$~will first run $M$ on $x$ producing some final
superposition of configurations $\sum_y \alpha_y \col{y}$ of $M$ on input $x$.
Then it will write a 0 or 1 in the extra track of the start cell of each
configuration, and run $M^R$ on this superposition
$\col{\phi} = \sum_y \alpha_y \col{y}\col{b_y}$.
If we were to instead run $M^R$ on the superposition
$\col{\phi'} = \sum_y \alpha_y \col{y}\col{M(x)}$ we would after $T(n)$ steps
have the superposition consisting entirely of the final configuration with
output $x;M(x)$. Clearly, $\iprod{\phi}{\phi'}$ is real, and since $M$ has
success probability at least $1 - \varepsilon/2$,
$\iprod{\phi}{\phi'} \geq \sqrt{1 - \varepsilon}$.
Therefore, since the time evolution of $M^R$ is unitary and hence
preserves the inner product, the final superposition of $M'$ must have
an inner product with $\col{x}\col{M(x)}$ which is real and at
least~$1 - \varepsilon/2$.  Therefore, the squared magnitude in the final
superposition of $M'$ of the final configuration with output~\mbox{$x;M(x)$}
must be at least $(1 - \varepsilon/2)^2 \geq 1 - \varepsilon$.
\end{proof}

\begin{corollary}
$\BQP^{\BQPs} = \BQP$.
\end{corollary}

\section*{Acknowledgement}

We wish to thank Bob Solovay for several useful discussions.

\samepage\pagebreak[1]

\end{document}